\documentclass[prb,citeautoscript,floatfix,twocolumn,amsmath,amssymb,superscriptaddress]{revtex4-1}%

\usepackage{graphicx}
\usepackage{dcolumn}
\usepackage{bm}
\usepackage{times}

\begin{document}

%\preprint{APS}

\title{Colloidal particle adsorption at liquid interfaces: 
Capillary driven dynamics\\ and thermally activated kinetics}
\author{Amir M. Rahmani}
\affiliation{Department of Mechanical Engineering, Stony Brook University, Stony Brook, NY 11794, USA.}

\author{Anna Wang}
\affiliation{Harvard John A. Paulson School of Engineering and Applied Sciences, Harvard University, Cambridge, MA 02138, USA.}

\author{Vinothan N. Manoharan}
\email[]{vnm@seas.harvard.edu}
\affiliation{Harvard John A. Paulson School of Engineering and Applied Sciences, Harvard University, Cambridge, MA 02138, USA.}
\affiliation{Department of Physics, Harvard University, Cambridge, MA 02138, USA.}

\author{Carlos E. Colosqui}
\email[]{carlos.colosqui@stonybrook.edu}
\affiliation{Department of Mechanical Engineering, Stony Brook University, Stony Brook, NY 11794, USA.}

%
%
%\date{\today}
%
%%%%%%%%%%%%%%%%%%%%%%%%%%%%%%%%%%%%%%%%%%%%%%%%%%%%%%%%%%%%%%%%%%%%%%%%%%%%%%%%%%%%%%%%%%%%%%%%%%%
%
% ABSTRACT
%
%%%%%%%%%%%%%%%%%%%%%%%%%%%%%%%%%%%%%%%%%%%%%%%%%%%%%%%%%%%%%%%%%%%%%%%%%%
%
\begin{abstract}
The adsorption of single colloidal microparticles (0.5--1 $\mu$m radius) at a water-oil interface has been recently studied experimentally using digital holographic microscopy [Kaz \textit{et al., Nat. Mater.}, 2012, \textbf{11}, 138--142]. 
An initially fast adsorption dynamics driven by capillary forces is followed by an unexpectedly slow relaxation to equilibrium that is logarithmic in time and can span hours or days.
The slow relaxation kinetics has been attributed to the presence of surface ``defects'' with nanoscale dimensions (1--5\,nm) that induce multiple metastable configurations of the contact line perimeter.
A kinetic model considering thermally activated transitions between such metastable configurations has been proposed [Colosqui \textit{et al., Phys. Rev. Lett.}, 2013, \textbf{111}, 028302] to predict both the relaxation rate and the crossover point to the slow logarithmic regime.
However, the adsorption dynamics observed experimentally before the crossover point has remained unstudied. 
In this work, we propose a Langevin model that is able to describe the entire adsorption process of single colloidal particles by considering metastable states produced by surface defects and thermal motion of the particle and liquid interface.
Invoking the fluctuation dissipation theorem, we introduce a drag term that considers significant dissipative forces induced by thermal fluctuations of the liquid interface. 
Langevin dynamics simulations based on the proposed adsorption model yield close agreement with experimental observations for different microparticles, capturing the crossover from (fast) capillary driven dynamics to (slow) thermally activated kinetics. 
\end{abstract}
%
%
%\pacs{}
%
%\keywords{}
\maketitle

%
%
%%%%%%%%%%%%%%%%%%%%%%%%%%%%%%%%%%%%%%%%%%%%%%%%%%%%%%%%%%%%%%%%%%%%%%%%%%%%%%%%%%%%%%%%%%%%%%%%%%
%   INTRODUCTION: NANOFLUIDICS - WETTING DYNAMICS -thermally activated PROCESSES
%%%%%%%%%%%%%%%%%%%%%%%%%%%%%%%%%%%%%%%%%%%%%%%%%%%%%%%%%%%%%%%%%%%%%%%%%%%%%%%%%%%%%%%%%%%%%%%%%%
%
\section{Introduction}
\footnotetext{\textit{$^{a}$~Department of Mechanical Engineering, Stony Brook University, Stony Brook, New York 11794, USA; E-mail: carlos.colosqui@stonybrook.edu}}
\footnotetext{\textit{$^{b}$~Harvard John A. Paulson School of Engineering and Applied Sciences, Harvard University, Cambridge, MA 02138, USA.}}
\footnotetext{\textit{$^{c}$~Department of Physics, Harvard University, Cambridge, MA 02138, USA. E-mail: vnm@seas.harvard.edu.}}

The adsorption and binding of colloidal particles at liquid interfaces is a fundamental phenomenon in colloid science that is relevant to applications in biomedical, environmental, food, and materials engineering.\cite{binks2006colloidal,kralchevsky2001}  
The ``irreversible'' binding of particles at liquid interfaces has been extensively exploited for emulsion stabilization,\cite{aveyard2003} encapsulation,\cite{dinsmore2002} wastewater treatment,\cite{rubio2002overview}, and self-assembly of novel materials.\cite{lin2003nanoparticle,biswas2008novel,mcgorty2010colloidal}   
Despite the success of diverse technological applications, many fundamental aspects of the adsorption of colloidal particles at liquid interfaces remain poorly understood.
Continuum thermodynamic descriptions are conventionally employed to predict the conditions under which colloidal particles bind to a liquid-fluid interface. 
Stable equilibrium conditions, where the system energy is minimized, are prescribed by the (Young) equilibrium contact angle determined by Young's law.
For colloidal particles larger than 10\,nm, the reduction of surface energy at equilibrium can be several orders of magnitude larger than the thermal energy, and thus binding to the interface can become ``irreversible''.\cite{pieranski1980two,aveyard1996}
While conventional thermodynamic models seem to describe certain fundamental aspects of equilibrium behavior, an increasing number of experimental observations highlight the necessity of developing accurate models for the (non-equilibrium) dynamics of adsorption.

A spontaneous and rapid relaxation to thermodynamic equilibrium is expected for perfectly spherical particles of nano- and microscale dimensions when driving forces associated with wetting and capillarity are much larger than damping forces due to hydrodynamic effects.\cite{huh1971hydrodynamic,petkov1995measurement,danov2000viscous,singh2005fluid,fischer2006viscous}
Nevertheless, experimental studies show that colloidal particle adsorption is neither spontaneous nor fast, owing to the presence of significant energy barriers caused by different physicochemical effects. 
The formation of electric double layers at liquid-liquid and liquid-solid interfaces is a common effect that prevents adsorption and/or hinders the relaxation to equilibrium.
Electrostatic energy barriers can be suppressed through the addition of electrolyte salts (e.g., NaCl) in sufficiently large concentration so that surface charges are screened.\cite{ohshima1982accurate,mbamala2002effective}
However, even for weak electrostatic effects, external work must be often applied through hydrodynamic shear or sonication in order to observe significant adsorption rates. \cite{stocco2011aqueous,du2010adsorption,garbin2012nanoparticles,Hongzhi2012}
Recent experimental work by Kaz {\it et al.} \cite{kaz2012} has studied the adsorption of single polystyrene microspheres with different surface functionalities at an interface between a glycerol/water mixture and decane.
Optical measurements by 3D digital holographic microscopy revealed a ``rapid'' initial adsorption of the particles following the spontaneous breach of the water-oil interface, which occurs when the salt concentration is sufficiently high ($>$100 mM NaCl).\cite{kaz2012}
The initial rapid adsorption is followed by a surprisingly slow relaxation to equilibrium that is logarithmic in time and has been attributed to thermally activated pinning/depinning of the contact line at nanoscopic surface defects with areas ranging from 1 to 30\,nm$^2$.\cite{kaz2012}  
Notably, the observed logarithmic relaxation can take place over several hours or even days and is reminiscent of physical aging in glassy systems with a free energy landscape that is densely populated by multiple local minima.\cite{struik1977physical,palmer1984models} 

The logarithmic relaxation reported by Kaz {\it et al.} \cite{kaz2012,wang2013} suggests that the interfacial energy of a colloidal particle straddling a liquid-liquid interface has multiple local minima induced by nanoscale surface defects. 
The interplay between nanoscale surface defects and thermal fluctuations can induce the thermally activated displacement of the three-phase contact line, which results in nontrivial wetting processes.\cite{prevost1999thermally,davitt2013thermally,colosqui2015}
Previous work by Colosqui {\it et al.} \cite{colosqui2013,razavi2014} proposed a kinetic model employing Kramers' theory of thermally activated transitions on a one-dimensional energy profile that has multiple minima induced by periodic energy barriers with magnitude $\Delta {\cal F}\propto A_d$ and period $\lambda = A_d/2\pi R$; here, $A_d$ is the characteristic defect area and $R$ is the particle radius.
The proposed analytical model predicts that the defect area $A_d$ and particle radius $R$ prescribe both the logarithmic relaxation rate and the critical distance from equilibrium where the adsorption process becomes logarithmic in time.

While previous work\cite{kaz2012,colosqui2013} has focused on rationalizing the slow logarithmic relaxation observed near equilibrium, no analytical models have been proposed to describe the observed adsorption dynamics preceding the slow logarithmic regime.
Recent experimental measurements of the (in-plane) diffusivity of microparticles at interfaces by Boniello {\it et al.}\cite{boniello2015} revealed that hydrodynamic drag models cannot explain the dissipative forces experienced by a particle straddling a liquid-fluid interface at different contact angles.  
Boniello {\it et al.} proposed that thermal motion of the liquid interface can induce a dominant contribution to (in-plane) dissipative forces that can be estimated via the fluctuation-dissipation theorem.\cite{kubo1966fluctuation}
Building on prior work by Colosqui {\it et al.},\cite{colosqui2013} this work introduces a Langevin model that is able to describe the entire adsorption dynamics, comprising (fast) capillary-driven and (slow) thermally activated regimes, by considering (1) a surface energy landscape with metastable states induced by nanoscale defects, (2) random fluctuations of surface forces induced by thermal motion of the colloidal particle and liquid-liquid interface, and (3) dissipative forces required to satisfy the fluctuation-dissipation relation.

%
%
%%%%%%%%%%%%%%%%%%%%%%%%%%%%%%%%%%%%%%%%%%%%%%%%%%%%%
%
% Langevin model for adsoprtion dynamics
%
%%%%%%%%%%%%%%%%%%%%%%%%%%%%%%%%%%%%%%%%%%%%%%%%%%%%%
\section{Theoretical model}
%
%\subsection{Langevin dynamics}
%
To model the adsorption dynamics we will assume that the particle undergoes uncorrelated Brownian motion. Therefore we employ a classical Langevin equation\cite{kubo1966fluctuation,hinch1975,colosqui2013} for the evolution of the (center-of-mass) vertical position $z(t)$.
For convenience, $z(t)$ is measured relative to the liquid-fluid interface located at $z=0$ (see Fig.~\ref{fig1}(a)).
Neglecting electrostatic and gravitational effects, the resulting force on the particle 
$F=F_s+F_d+F_{t}$ 
is the sum of surface forces $F_s$,
dissipative forces $F_d$, 
and forces due to thermal motion $F_{t}$.
We will consider that the surface force $F_s=\langle F_s \rangle+F'_s$ comprises a mean deterministic component $\langle F_s \rangle=-\partial {\cal F}/\partial z$ determined by the surface free energy ${\cal F}$, and a zero-mean random force $F'_s$ due to thermal fluctuations of the liquid interface. 
The fluctuation-dissipation theorem \cite{kubo1966fluctuation,hinch1975} states that dissipative forces $F_d=-\xi\dot{z}$ and random thermal forces $F_{t}=\sqrt{2k_B T \xi} \eta$ are related through the effective friction coefficient $\xi$; here, $\eta(t)$ is a (zero-mean unit-variance) white noise, $k_B$ is the Boltzmann constant and $T$ is the temperature of the fluids.
The proposed Langevin equation for the particle adsorption dynamics thus reads\cite{colosqui2013}
\begin{equation}
m \ddot{z}=-\frac{\partial {\cal F}}{\partial z}-\xi\dot{z}+\sqrt{2 k_B T \xi} \eta(t), 
\label{eq:langevin}
\end{equation}
where $m$ is the particle mass. 
The Langevin model in eqn~(\ref{eq:langevin}) is valid when the particle is in thermal equilibrium with the surrounding fluids and its position is subject to uncorrelated Brownian motion.
Despite the simplifying assumptions, eqn~(\ref{eq:langevin}) has been able to produce close agreement with fully atomistic molecular dynamics simulations in previous work\cite{colosqui2013} where the fluctuations of the interface and particle may be correlated.
%

%%%%%%%%%%%%%%%%%%%%%%%%%%%%%%%%%%%%%%%%%%%%%%%%%%%%%%%%%%%
% THEORY: SURFACE FORCES SPATIAL AND THERMAL FLUCTUATIONS
%%%%%%%%%%%%%%%%%%%%%%%%%%%%%%%%%%%%%%%%%%%%%%%%%%%%%%%%%%%
\subsection{Surface forces}
\begin{figure}
\center
\includegraphics[angle=0,width=.95\linewidth]{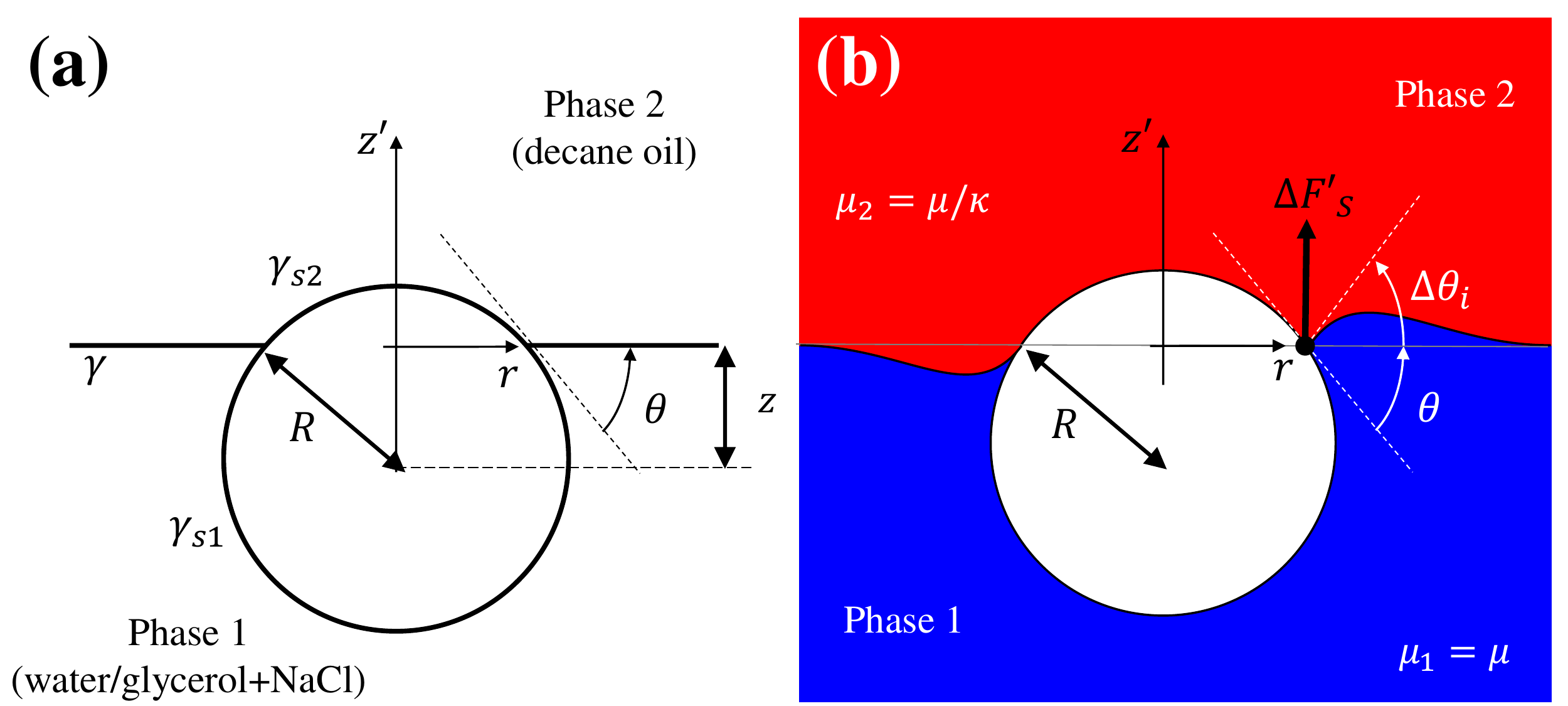}
\caption{Colloidal particle at an interface.
(a) Spherical particle of radius $R$ straddling a sharp flat interface between two immiscible phases at a contact angle $\theta=\arccos(-z/R)$.
The surface energies $\gamma$, $\gamma_{s1}$, $\gamma_{s2}$ determine the expected equilibrium position $z_E=-R\cos\theta_E$, where $\cos\theta_E=(\gamma_{s2}-\gamma_{s1})/\gamma$. 
For nanoscale surface defects ($\sqrt{A_d}\ll R$), the free energy profile ${\cal F}(z)$ is given by eqn~(\ref{eq:energy}). 
(b) Fluctuating surface forces $F_s'=\sum_i \Delta F_i'$ along the $z$-direction are induced by thermal fluctuations of the local contact angle $\theta_i=\theta + \Delta \theta_i$.
The total surface force $F_s=-\partial{\cal F}/\partial z + F_s'$ comprises average (deterministic) and  fluctuating (random) components.
}
\label{fig1}
\end{figure}

Within the framework of continuum thermodynamics, changes in the Helmholtz free energy of a colloidal particle at a liquid-fluid interface give the reversible work performed by the system at constant volume and temperature.
If we assume symmetry about the polar and azimuth axes, the surface free energy ${\cal F}={\cal F}(z)$ can be parametrized with respect to the particle center of mass position $z$, measured normal to the flat liquid-fluid interface (see Fig.~\ref{fig1}a). 
The surface free energy of a spherical particle of radius $R$ straddling a perfectly flat liquid interface located at $z=0$ can be expressed as \cite{aveyard1996,colosqui2013} 
\begin{equation}
{\cal F}_S(z)=\gamma \pi (z-z_E)^2+ \tau 2\pi \sqrt{R^2-z^2}+C
\label{eq:energy_smooth} 
\end{equation}
for $|z|\le R$, where $z_E=-R \cos\theta_E$ is the equilibrium position determined by the (Young) equilibrium contact angle $\theta_E$, $\gamma$ is the liquid-fluid surface tension, $\tau$ is the line tension, and $C$ is an additive constant. 

While the surface tension $\gamma$ is always positive and has values on the order of 10 mN/m for water-oil interfaces, the line tension $\tau$, associated with the excess free energy at the three-phase contact line, can be either positive or negative with a magnitude on the order of 10 pN, according to theoretical estimates.\cite{amirfazli2004status}
It can be challenging to determine the line tension empirically. Different experimental studies with simple liquids on smooth substrates report values of $\tau$ ranging between 1 and 1000 pN. \cite{widom1995line,drelich1996significance,checco2003nonlinear,berg2010impact}  
Defining a dimensionless line tension $\bar{\tau}=\tau/\gamma R$, we can identify two situations in which forces induced by line tension cannot be neglected: (1) when the particle barely straddles the liquid interface, such that $|z-R|<\bar{\tau}^2/2(1+\cos\theta_E)^2$, and (2) when the particle is very close to equilibrium, such that $|z-z_E|<\bar{\tau}/\tan\theta_E$.  
For the microparticles and water-oil interface studied in this work\cite{kaz2012} we have $R \simeq 1$~$\mu$m, $\gamma \simeq 4\times 10^{-2}$~N/m and $\theta_E=110^\circ$, which gives $\bar{\tau}\lesssim$~$10^{-4}$--$10^{-1}$. Thus, the line tension contributions in eqn~(\ref{eq:energy_smooth}) can be neglected for the experimentally observed trajectories $z(t)$.

For perfectly spherical particles and negligible line tension contributions, eqn~(\ref{eq:energy_smooth}) predicts a single energy minimum at $z=z_E$ and a relaxation to equilibrium driven by  monotonic reduction of surface energy.
According to eqn~(\ref{eq:energy_smooth}), when colloidal particles with nano- or microscale dimensions ($R>10$\,nm) approach equilibrium $z\to z_E$, they become ``irreversibly'' adsorbed at the 
interface because the depth of the energy reduction (i.e., the binding energy) is several orders of magnitude larger than the thermal energy.
%

%%%%%%%%%%%%%%%%%%%%%%%%%%%%%%%%%%%%%%%%%%%%%%%%%%%%%%%%%%%%%%%%%%%%%%%%%%%%%%%%%%%%%%%%%%%%%%%%%%%%%%%%%%%%%%%%%%%%%%%%%%%%%%%%
% Figure1: Problem setup
%%%%%%%%%%%%%%%%%%%%%%%%%%%%%%%%%%%%%%%%%%%%%%%%%%%%%%%%%%%%%%%%%%%%%%%%%%%%%%%%%%%%%%%%%%%%%%%%%%%%%%%%%%%%%%%%%%%%%%%%%%%%%%%%
%%%%%%%%%%%%%%%%%%%%%%%%%%%%%%%%%%%%%%%%%%%%%%%%%%%%%%%%%%%%%%%%%%%%%%%%%%%%%%%%%%%%%%%%%%%%%%%%%%%%%%%%%%%%%%%%%%%%%%%%%%%%%%%%
%

\subsubsection{Nanoscale surface defects.~~}
Experimental observations for colloidal microparticles\cite{kaz2012,wang2013} have revealed an unexpectedly slow approach to equilibrium due to the presence of nanoscale surface defects not considered in eqn~(\ref{eq:energy_smooth}).
As in previous work by Colosqui {\it et al.} \cite{colosqui2013} we aim to develop an analytical model that incorporates local minima in the surface energy profile ${\cal F}(z)$ that are induced by multiple surface defects with nanoscale dimensions.
For this purpose, we assume that the surface of the particle is densely populated by defects with a characteristic nanoscale area $A_d\sim{\cal O}(1\, \mathrm{nm}^2)$ that produce surface energy perturbations of magnitude $\Delta {\cal F} \simeq\gamma A_d > k_B T$.
The displacement of the contact line over a single defect of characteristic area $A_d$ changes the surface energy by an amount $\Delta {\cal F}\sim\gamma A_d$ and the average contact line position by an amount $\lambda = A_d/2\pi R\ll R$.
To model this process we introduce a periodic spatial perturbation of magnitude $\Delta {\cal F}$ and period $\lambda$ in the particle free energy so that
\begin{equation}
{\cal F}(z)={\cal F}_S(z)
+ \frac{1}{2} \Delta {\cal F} \sin\left(\frac{2 \pi}{\lambda}(z-z_E)-\varphi\right).
\label{eq:energy}
\end{equation}
The phase $\varphi$ of the spatial perturbation produces a negligible shift $\Delta z_E\le \lambda \ll R$ of the equilibrium position and thus can be arbitrarily chosen; we adopt $\varphi=-\pi/2$ so that the global energy minimum in eqn~(\ref{eq:energy}) remains at the equilibrium position $z=z_E$ determined by the equilibrium contact angle $\theta_E$.

\subsubsection{Fluctuating surface forces.~~}
For the case of a liquid-fluid interface fluctuating due to thermal motion, the surface force can be decomposed into mean and fluctuating components 
$F_s=\langle F_s \rangle+F'_s$, where
$\langle F_s \rangle=-\partial {\cal F}/\partial z$ and 
$\langle F'_s \rangle=0$; hereafter, angle brackets $\langle ~ \rangle$ indicate ensemble averages.
The effects of spatial fluctuations of the contact line induced by localized defects on the particle surface are considered in the free energy profile ${\cal F}(z)$ in eqn~(\ref{eq:energy}), and thus in the deterministic mean surface force $\langle F_s \rangle$. 
We will consider the case when the fluctuating force $F'_s$ is due to thermal motion of the liquid-fluid interface, which causes random fluctuations of the local contact angle $\theta_i=\langle \theta_i\rangle + \Delta \theta_i$, where the average contact angle is $\langle \theta_i\rangle=\theta=\arccos(-z/R)$ (see Fig.~\ref{fig1}b) and $\langle \Delta \theta_i \rangle=0$.

The fluctuating surface force normal to the interface can be estimated by summing small contributions $\Delta F'_s\simeq\gamma s_i (\cos\theta_i-\cos\theta)/\sin\theta$  (see Fig.~\ref{fig1}b) from contact line segments having nanoscale dimensions $s_i \simeq \sqrt{A_d}$. 
In similar fashion, Boniello {\it et al.} have estimated fluctuating surface forces parallel to the interface.\cite{boniello2015}
Assuming small random fluctuations $|\Delta \theta_i| < 1$ of the local contact angle, we obtain the fluctuating surface force
\begin{equation}
F'_s \simeq \gamma \sqrt{A_d} \sum_{i=1}^{N} \Delta \theta_i,
%\frac{\cos(\theta+\Delta\theta_i)-\cos(\theta)}{\sin\theta} \simeq \gamma l .
\label{eq:f's}
\end{equation}
where $N=2\pi R \sin\theta/\sqrt{A_d}$.
The fluctuating surface force $F'_s$ given by eqn~(\ref{eq:f's}) has zero mean $\langle F'_s \rangle=0$ but finite variance $\langle {F_s'}^2\rangle>0$, and thus produces a finite energy input to the translational motion of the particle.

%%%%%%%%%%%%%%%%%%%%%%%%%%%%%%%%%%%%%%%%%%%%%%%%%%%%
% THEORY: DISSIPATIVE FORCES
%%%%%%%%%%%%%%%%%%%%%%%%%%%%%%%%%%%%%%%%%%%%%%%%%%%%
\subsection{Dissipative forces}
A key aspect of the Langevin dynamics modeled by eqn~(\ref{eq:langevin}) is that deterministic dissipative forces and random forces due to thermal fluctuations are related through the effective friction coefficient $\xi$, in accordance with the fluctuation-dissipation theorem. 
For systems in thermal equilibrium, the fluctuation-dissipation relation ensures that the energy input due to random thermal forces is strictly balanced, in the ``long-time'' limit, by dissipation of energy to the surrounding fluids.\cite{kubo1966fluctuation,hinch1975}    
The friction coefficient $\xi$ can be estimated from different theoretical models for predicting dissipative forces.
Alternatively, it is possible to determine the effective friction coefficient by employing the Green--Kubo relation\cite{kubo1966fluctuation,searles2000}  
$\xi=(1/k_B T)\int_{0}^\infty \langle F_{t}(0) F_{t}(t) \rangle dt$,
which involves the time autocorrelation of the random thermal forces $F_{t}$ acting on the particle. 
We will assume that the thermal force $F_{t}=F'_{z}+F'_{s}$ normal to the liquid interface is the sum of the random translational force $F'_z$ due to thermal fluctuations of the fluid bulk and the random surface force $F'_s$ due to thermal fluctuations of the liquid interface.
Further assuming that these random forces are uncorrelated $(\langle F'_{z} F'_{s}\rangle=0)$, the friction coefficient $\xi=\xi_z+\xi_s$ is the sum of $\xi_z$, which will be determined by dynamic arguments, and $\xi_s$, which will be estimated by the fluctuation-dissipation relation as proposed recently by Boniello {\it et al.}\cite{boniello2015}
\subsubsection{Translational drag: hydrodynamics and contact line dynamics.~~}
Under creeping flow conditions the friction coefficient associated with translational motion can be generally expressed as 
$\xi_z(z)=(\mu_1 K_1 + \mu_2 K_2) R =(K_1+K_2/\kappa) \mu R $, 
where $\mu_{1}=\mu$ and $\mu_{2}=\mu/\kappa$ are the viscosities of each fluid phase, $\kappa=\mu_1/\mu_2>1$ is the viscosity ratio,  
and $K_{1,2}$ are position-dependent drag coefficients that can be estimated using available hydrodynamic models.\cite{huh1971hydrodynamic,petkov1995measurement,danov2000viscous,singh2005fluid,fischer2006viscous}
Given the relatively large viscosity ratio $\kappa=12.8$ for the experimental conditions studied in this work\cite{kaz2012}, $\mu=\mu_1$ is adopted hereafter as the characteristic viscosity.  
As in previous studies,\cite{kaz2012,boniello2015} we find that translational friction coefficients $\xi_z$ predicted by different hydrodynamic models for creeping flow conditions\cite{huh1971hydrodynamic,petkov1995measurement,danov2000viscous,singh2005fluid,fischer2006viscous} are two to three orders of magnitude smaller than those required to agree with experimental observations [see Sec.~\ref{sec:drag} for details].

The translational motion of the contact line perimeter can be considered to produce dissipation of energy. 
According to the molecular kinetic theory (MKT) developed by Blake and co-workers\cite{blake1969kinetics,blake2006physics,de1999dynamic}, the energy ``consumed'' in the adsorption and desorption of molecules at the moving contact line can be modeled as a virtual dissipative force 
$F_d=-\xi_{z} \dot{z}$ determined by the friction coefficient
$\xi_{z}=2\pi R \sin \theta \mu (\nu/\lambda_{M}^3)\exp(W_a/k_B T)$, 
where $\nu$ is the molecular volume of the liquid in phase 1, 
$W_a=\gamma\lambda_{M}^2(1+\cos\theta_E)$ is the work of adhesion,
and $\lambda_{M}$ is the characteristic size of the adsorption sites.
The characteristic size  $\lambda_{M}$ of the adsorption sites is treated as a free parameter that must, however, remain within molecular dimensions (i.e., $\lambda_{M}<$~1\,nm) for consistency with the premises of MKT.\cite{ramiasa2014}  
As discussed in detail later in Sec.~\ref{sec:drag}, maximum friction coefficients 
$\xi_s(z=0)={\cal O}$(0.1--1\,$\mu$N\,s/m) predicted by MKT for $\lambda_{M}=$~0.5--1\,nm produce dissipative forces that are two to three orders of magnitude smaller than those required to describe the experimentally observed adsorption dynamics. 
As reported in previous experimental studies,\cite{kaz2012,boniello2015} we find that damping forces due to hydrodynamics and molecular kinetic effects considered by MKT cannot fully explain the dynamic behavior of particles at interfaces.%

\subsubsection{Interfacial fluctuations: Green--Kubo relations.~~}
The thermal motion of the liquid interface causes fluctuating surface forces on the particle (cf. Fig.~\ref{fig1}b) that must be balanced by dissipative forces $F_d=-\xi_s \dot{z}$ in order to satisfy the fluctuation-dissipation relation. 
Hence, the friction coefficient $\xi_s$ can be estimated by the Green--Kubo relation
\begin{eqnarray}
\xi_s=\frac{1}{k_B T}\int_{0}^{\infty} \langle F'_{s}(0) F'_{s}(t) \rangle dt
\simeq \frac{1}{k_B T} \langle {F'_{s}}^2 \rangle \times t_c,
\label{eq:green-kubo}
\end{eqnarray}
where $t_c$ is the correlation time, assuming an Ornstein--Uhlenbeck process for the relaxation of the interface.
In order to estimate the friction coefficient in eqn~(\ref{eq:green-kubo}) we will begin by considering that contact angle fluctuations $\Delta\theta_i$ along the contact line perimeter relax through capillary waves that travel at a speed $v_c=\gamma/\mu$. 
We can then estimate the correlation time $t_c=\chi (2\pi R \sin\theta/v_c)$ where $\chi={\cal O}(1)$ is a scaling factor accounting for the number of times capillary waves travel around the contact line perimeter before correlation between local fluctuations is completely lost.

Invoking eqn~(\ref{eq:f's}) the variance of the fluctuating force is 
$\langle {F_s'}^2\rangle =\gamma^2 \sqrt{A_d} (2\pi R \sin\theta) \langle\Delta\theta_i^2\rangle$,
where the mean square fluctuation of the local contact angle $\langle\Delta\theta_i^2\rangle \simeq k_B T/ \gamma A_d$ is determined by a balance between thermal energy and elastic surface deformation.
Incorporating the expressions above into eqn~(\ref{eq:green-kubo}) we arrive at the simple expression
\begin{eqnarray}
\xi_s = \chi \mu \frac{4\pi^2 R^2}{\sqrt{A_d}} \left(1-(z/R)^2\right)
\label{eq:xis}
\end{eqnarray}
for the friction coefficient determining dissipative forces induced by thermal fluctuations of the interface.
For the case of microparticles $\left(R={\cal O}(1~\mu\mathrm{m})\right)$ with nanoscale defect areas 
$\left(A_d={\cal O}(1~\mathrm{nm}^2)\right)$, eqn~(\ref{eq:xis}) predicts a maximum friction coefficient 
$\xi_s(z=0)={\cal O}(100~\mu\mathrm{N~s/m})$, which has the magnitude required for agreement between the Langevin model in eqn~(\ref{eq:langevin}) and experimental observations.\cite{kaz2012,wang2013}

%%%%%%%%%%%%%%%%%%%%%%%%%%%%%%%%%%%%%%%%%%%%%%%%%%%%%%%%%%%%%%%%%%%%%%%%%%
%  THEORY: CROSSOVER
%%%%%%%%%%%%%%%%%%%%%%%%%%%%%%%%%%%%%%%%%%%%%%%%%%%%%%%%%%%%%%%%%%%%%%%%%%
\subsection{Crossover to thermally activated adsorption}
As the particle approaches equilibrium, where $2\gamma|z- z_E|< \Delta {\cal F}/\lambda$, the free energy ${\cal F}$ in eqn~(\ref{eq:energy}) has local energy minima where $\partial {\cal F}/\partial z = 0$ and the mean surface force vanishes. 
As a result, sufficiently close to equilibrium the adsorption dynamics described by eqn~(\ref{eq:langevin}) is no longer driven by a monotonic reduction of surface energy and becomes dominated by thermally activated transitions between multiple metastable states.\cite{colosqui2013}
The crossover to thermally activated adsorption is a gradual process that takes place over a finite position interval $|z-z_C|>0$ that is approximately centered at the crossover point $z(t)=z_C$. 
According to previous work by Colosqui {\it et al.}\cite{colosqui2013} the crossover point $z_C$ can be estimated by the equation
\begin{equation}
|z_C-z_E|=\alpha \frac{\Delta{\cal F}}{2 \gamma \lambda}
\label{eq:crossover}
\end{equation}
for $0.5\lesssim\alpha\lesssim 1$.
The crossover position $z_C$ is prescribed by the free energy profile and is independent of frictional and dissipative effects.
Provided that both the energy barrier $\Delta{\cal F}$ and period $\lambda$ employed in eqns~(\ref{eq:langevin})--(\ref{eq:energy}) are properly determined, the crossover criterion in eqn~(\ref{eq:crossover}) can estimate the critical distance from equilibrium below which the adsorption becomes logarithmic in time. 
%

%%%%%%%%%%%%%%%%%%%%%%%%%%%%%%%%%%%%%%%%%%%%%%%%%%
%
%	RESULTS
%
%%%%%%%%%%%%%%%%%%%%%%%%%%%%%%%%%%%%%%%%%%%%%%%%%%
\section{Results}
The proposed adsorption model consisting of eqns~(\ref{eq:langevin})--(\ref{eq:energy}) and (\ref{eq:xis}) has formally three adjustable parameters: (i) the energy barrier magnitude $\Delta {\cal F}$, and (ii) the period 
$\lambda$ prescribing the free energy profile in eqn~(\ref{eq:energy}), along with (iii) the dimensionless correlation time, or correlation factor, $\chi\sim{\cal O}(1)$ determining the friction coefficient in eqn~(\ref{eq:xis}).
However, the energy barrier magnitude and period must be determined by the same nanoscale defect area $A_d$ for consistency with physical arguments. 
Employing a correlation factor $\chi=3$ for the friction coefficient in eqn~(\ref{eq:xis}) produces the dynamics reported in Figs.~\ref{fig2}--\ref{fig4}. 
While small variations in the energy barrier $\Delta {\cal F}$ and period $\lambda$ produce large differences in the modeled dynamics, the correlation factor $\chi$ can be varied by about 20\% without observing significant effects.

For the functionalized polystyrene microspheres ($R\simeq$~1~$\mu$m) studied in this work, previous experiments\cite{kaz2012,wang2013} have found that the defect area $A_d$ ranges from 5 to 30 nm$^2$ and is uncorrelated with the particle radius.
The experimental system is at room temperature ($T=21\pm 3^\circ$C) and consists of two immiscible phases: phase 1 is a NaCl solution ($>$100 mM NaCl) in a water/glycerol (45:55\% w/w) mixture with viscosity 11.5~mPa~s, and phase 2 is a anhydrous decane with viscosity 0.9~mPa~s. 
The viscosity $\mu=11.5$~mPa~s of the aqueous phase (phase 1) is adopted as the characteristic viscosity given the large viscosity ratio ($\kappa>10$).
The surface tension at the water-oil interface is $\gamma=0.037$\,N/m and the equilibrium contact angle for all the studied particles is $\theta_E=110^\circ$.\cite{kaz2012}
The experimental measurements have a time resolution of 0.2--10\,ms and spatial resolution of 1\,nm.

The energy barrier induced by a (three-dimensional) surface defect on the particle surface involves a change of (1) the liquid-liquid interface surface area $\Delta A_{12}$ and (2) the solid wetted area $\Delta A_{s1}$, both comparable but not necessarily equal to characteristic defect area $A_d$.\cite{colosqui2013,razavi2014} 
Hence, the magnitude of the energy barrier $\Delta{\cal F}$ in eqn~(\ref{eq:energy}) is given by 
$\Delta{\cal F}\simeq |\gamma (\Delta A_{12} + \Delta A_{s1}\cos\theta_d)| = \beta \gamma  A_d$ where $\beta\sim{\cal O}(1)$ is a shape factor prescribed by the specific geometry of the defect and the Young equilibrium contact angle $\theta_d$ determined by local surface energies.
%induced by wetting/dewetting a defect on the particle surface
%
Naively assuming homogeneous surface energy so that $\theta_d=\theta_E$ and a spherical cap defect with projected area $A_d$ and protruding height $h_d=$~0--0.5$\sqrt{A_d/\pi}$ leads to $\beta=$~0.35--0.2, which is close to the values employed for fitting the experimental results [see Table~\ref{tb:parameters}].

\begin{table*}
\small
  \caption{\ Model parameters employed for Langevin simulations (eqns~(\ref{eq:langevin})--(\ref{eq:energy})) of colloidal particles experimentally studied in Ref.~\citenum{kaz2012}.}
  \label{tb:parameters}
  \begin{tabular*}{\textwidth}{@{\extracolsep{\fill}}lcccc}
    \hline
    Particle class & $A_d$ (nm$^2$)& $\Delta {\cal F}=\beta \gamma A_d$ & $\lambda=A_d/2\pi R$ (pm) & $\beta$ \\
    \hline
    Carboxyl & 9.2 & 19.0 $k_B T$ & 1.63 & 0.23 \\
    Sulphate & 12.9 & 48.9 $k_B T$ & 2.34 & 0.41 \\
    Carboxylate-modified  & 85.7 & 360 $k_B T$ & 15.5 & 0.45\\
    \hline
  \end{tabular*}
\end{table*}

\subsection{Langevin simulations}
%

%%%%%%%%%%%%%%%%%%%%%%%%%%%%%%%%%%%%%%%%%%%%%%%%%%%%%%%%%%%%%%%%%%%%%%%%%%%%%%%%%%%%%%%%%%%%%%%%%%%%%%%%%%%%%%%%%%%%%%%%%%%%%%%%
% Figure1: Problem setup
%%%%%%%%%%%%%%%%%%%%%%%%%%%%%%%%%%%%%%%%%%%%%%%%%%%%%%%%%%%%%%%%%%%%%%%%%%%%%%%%%%%%%%%%%%%%%%%%%%%%%%%%%%%%%%%%%%%%%%%%%%%%%%%%
\begin{figure}
\centering
\includegraphics[angle=0,width=.95\linewidth]{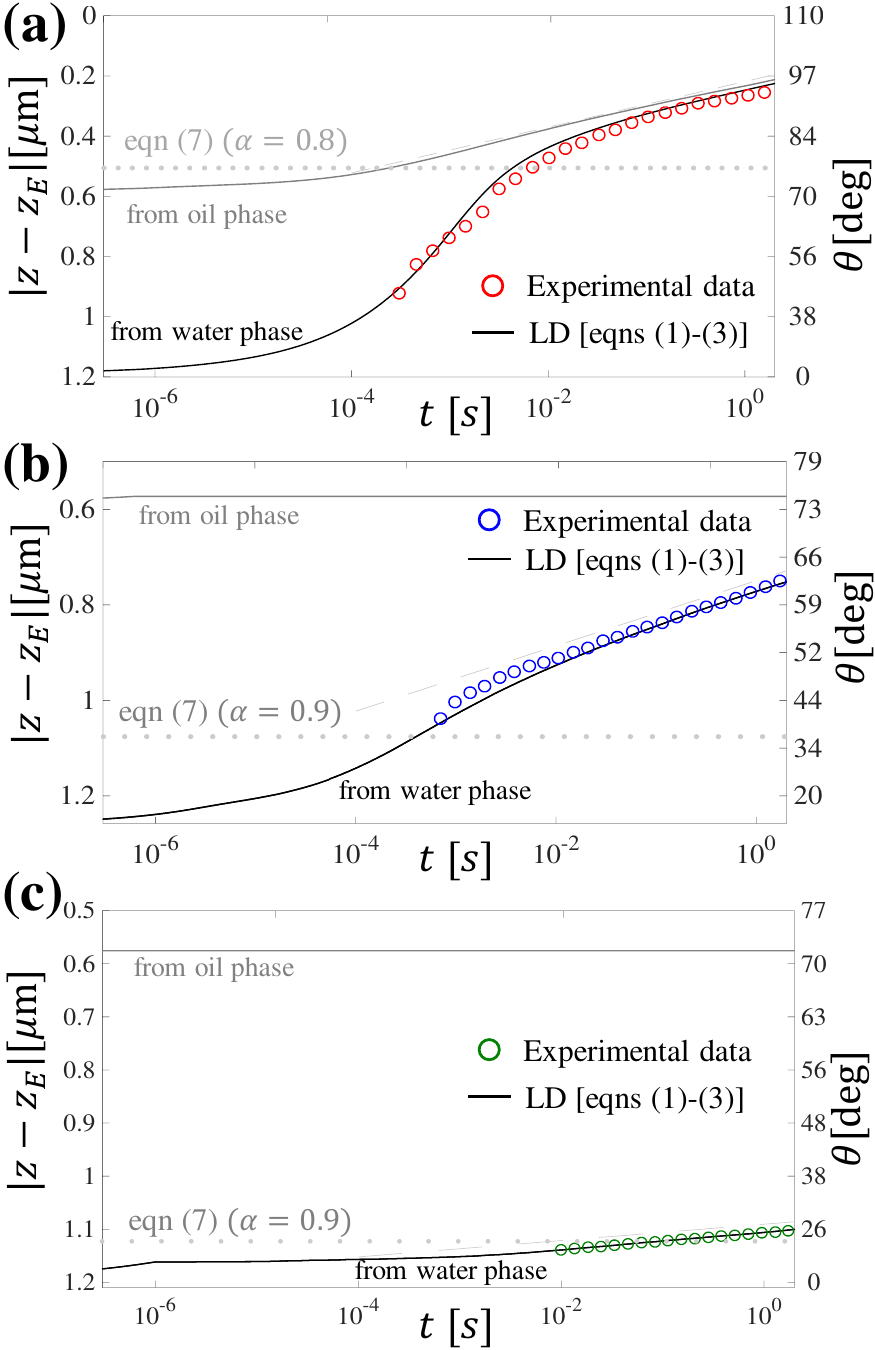}
\vskip -5pt
\caption{Adsorption dynamics: Langevin dynamics versus experiments.
Separation from equilibrium $|z(t)-z_E|$ versus time for three polystyrene particles of radius $R=0.9\,\mu\mathrm{m}$ with different surface functional groups. (a) Carboxyl, (b) Sulphate, and (c) Carboxylate-modified.\cite{kaz2012}
The contact angle on the right axis corresponds to $\theta=\arccos(-z/R).$
Markers: Experimental observations from Ref.~\citenum{kaz2012}.
Solid lines: LD simulation (eqns~(\ref{eq:langevin})--(\ref{eq:energy})) using model parameters reported in Table~\ref{tb:parameters}; the adsorption is initiated from the water and oil phase.
Dashed lines: Logarithmic-in-time fits $|z(t)-z_E|=c_1+c_2\log t$ ($c_{1,2}=$~const.).
Horizontal dotted lines: Crossover distance $|z_C-z_E|$  determined by eqn~(\ref{eq:crossover}) for $\alpha=$~0.8--0.9.
}
\vskip -5pt
\label{fig2}
\end{figure}
%%%%%%%%%%%%%%%%%%%%%%%%%%%%%%%%%%%%%%%%%%%%%%%%%%%%%%%%%%%%%%%%%%%%%%%%%%%%%%%%%%%%%%%%%%%%%%%%%%%%%%%%%%%%%%%%%%%%%%%%%%%%%%%%

In Table~\ref{tb:parameters} we report the model parameters employed in eqns~(\ref{eq:langevin})--(\ref{eq:energy}) for the Langevin Dynamics (LD) simulations in this work.
The LD simulations must employ extremely small time steps $\Delta t= 0.1$~ps to accurately resolve forces produced by the free energy in eqn~(\ref{eq:energy}), which has spatial oscillations with very small periods 
$\lambda=A_d/2\pi R=$~1.6--3.6~pm [see Table~\ref{tb:parameters}]. 
Simulated particle trajectories $z(t)$ reported in Fig.~\ref{fig2} for three different functionalized particles (carboxyl, sulphate, and carboxylate-modified surface groups) of radius $R=0.9$~$\mu$m are in close agreement with experimental observations for adsorption initiated from the aqueous phase.\cite{kaz2012}   

Notably, LD simulations describe both the initial adsorption dynamics, driven by capillary forces induced by reduction of surface energy, and the transition to a thermally activated regime where the particle follows a logarithmic-in-time trajectory as predicted in Ref.~\citenum{colosqui2013}.  
The crossover position $z_C$ in all studied cases can be estimated by the criterion in eqn~(\ref{eq:crossover}) using $\alpha=$~0.8--0.9 [see Fig.~\ref{fig2}]. 
Simulations for three studied particles show that the adsorption dynamics can be dramatically different when it is initiated from the oil phase, where the initial position $z(t=0)$ is much closer to equilibrium. 
Extremely low relaxation rates and kinetically trapped states are observed for large energy barriers $\Delta {\cal F}>20~k_B T$ [cf. Fig.~\ref{fig2}b--c] when the adsorption is initiated in the region $|z(t=0)-z_E|<|z_C-z_E|$, and the entire process is dominated by thermally activated transitions between metastable states.

\subsection{Particle size effects}
\begin{figure}
\centering
\includegraphics[angle=0,width=.95\linewidth]{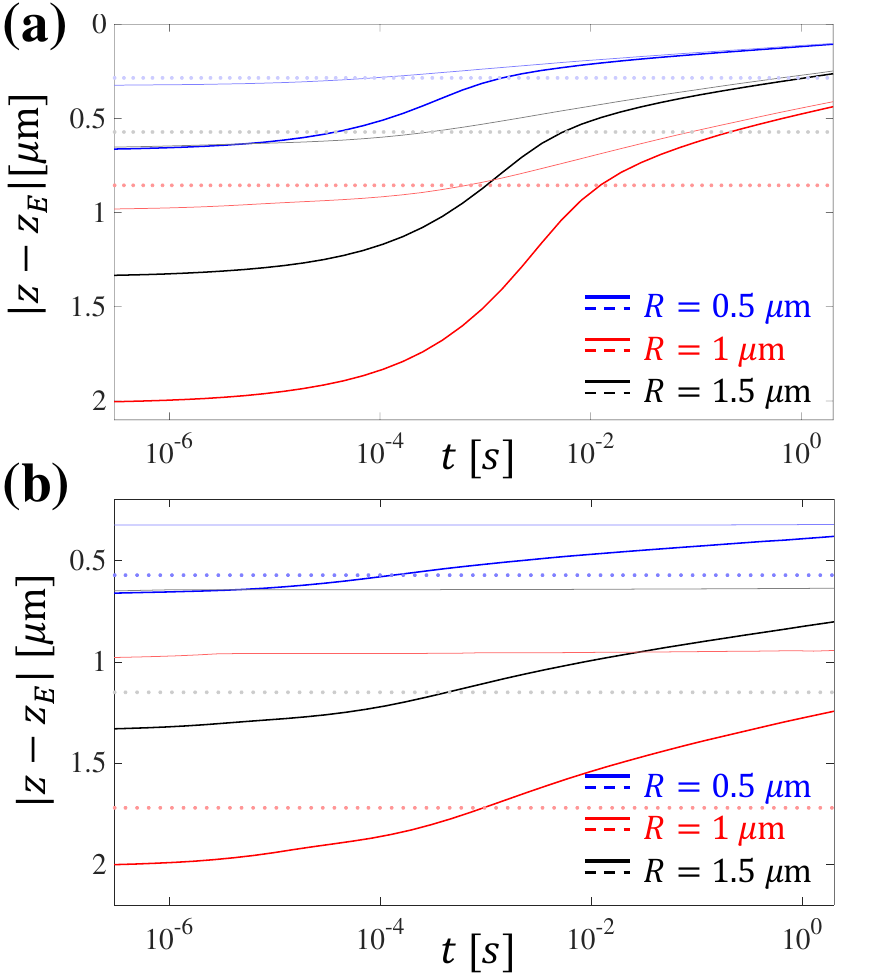}
\vskip -5pt
\caption{Particle size variation.
Separation from equilibrium $|z(t)-z_E|$ versus time for particles of radius $R=$~0.5, 1, and 1.5~$\mu\mathrm{m}$:  
(a) Carboxyl-functionalized particles, (b) Sulphate-functionalized particles.
Solid lines: LD simulations of eqns~(\ref{eq:langevin})--(\ref{eq:energy}) using parameters reported in Table~\ref{tb:parameters} for 
adsorption initiated from the water phase (thicker lines) and from the oil phase (thinner lines). 
Horizontal dotted lines: Crossover position predicted by eqn~(\ref{eq:crossover}) for $\alpha=$~0.8 (a) and $\alpha=$~0.9 (b).}
\label{fig3}
\vskip -5pt
\end{figure}

Experimental observations by Kaz {\it et al.}\cite{kaz2012} have reported that the defect area $A_d$ is uncorrelated with the particle size and that it is predominantly determined by the surface functionalization.
A series of LD simulations for particles with a fixed defect area ($A_d=$~9.2 nm$^2$ and 12.9 nm$^2$) and three different radii $R=$~0.5, 1, and 1.5~$\mu$m are reported in Fig.~\ref{fig3}.
Despite the fact that damping forces scale with $R^2$, the adsorption dynamics of smaller particles is not significantly faster.
When the radius of the particle is reduced the adsorption dynamics enters the slow logarithmic regime much earlier, which is consistent with the crossover criterion in eqn~(\ref{eq:crossover}).  
Simulations for the studied range of microparticle radii $R=$~0.5--1.5~$\mu$m with energy barriers much larger than the thermal energy $\Delta{\cal F}>20~k_B T$ (cf. sulphate functionalized particles in Fig.~\ref{fig3}(b)) reveal that the logarithmic-in-time relaxation to equilibrium cannot be observed within the experimental observation window (0 to 2~s) if the adsorption is initiated from the oil phase; these findings are consistent with recent experimental observations.\cite{wang2013} 

%  
%%%%%%%%%%%%%%%%%%%%%%%%%%%%%%%%%%%%%%%%%%%%%%%%%%%%%%%%%%%%%%%%%%%%%%%%%%%%%%%%%%%%%%%%%%%%%%%%%%%%%%%%%%%%%%%%%%%%%%%%%%%%%%%%
% Figure3: Size effects
%%%%%%%%%%%%%%%%%%%%%%%%%%%%%%%%%%%%%%%%%%%%%%%%%%%%%%%%%%%%%%%%%%%%%%%%%%%%%%%%%%%%%%%%%%%%%%%%%%%%%%%%%%%%%%%%%%%%%%%%%%%%%%%%
%%%%%%%%%%%%%%%%%%%%%%%%%%%%%%%%%%%%%%%%%%%%%%%%%%%%%%%%%%%%%%%%%%%%%%%%%%%%%%%%%%%%%%%%%%%%%%%%%%%%%%%%%%%%%%%%%%%%%%%%%%%%%%%%

\subsection{Dissipative effects: Drag models comparison \label{sec:drag}}
%

%%%%%%%%%%%%%%%%%%%%%%%%%%%%%%%%%%%%%%%%%%%%%%%%%%%%%%%%%%%%%%%%%%%%%%%%%%%%%%%%%%%%%%%%%%%%%%%%%%%%%%%%%%%%%%%%%%%%%%%%%%%%%%%%
% Figure4: Binding Energy
%%%%%%%%%%%%%%%%%%%%%%%%%%%%%%%%%%%%%%%%%%%%%%%%%%%%%%%%%%%%%%%%%%%%%%%%%%%%%%%%%%%%%%%%%%%%%%%%%%%%%%%%%%%%%%%%%%%%%%%%%%%%%%%%
\begin{figure}
\centering
\includegraphics[angle=0,width=.95\linewidth]{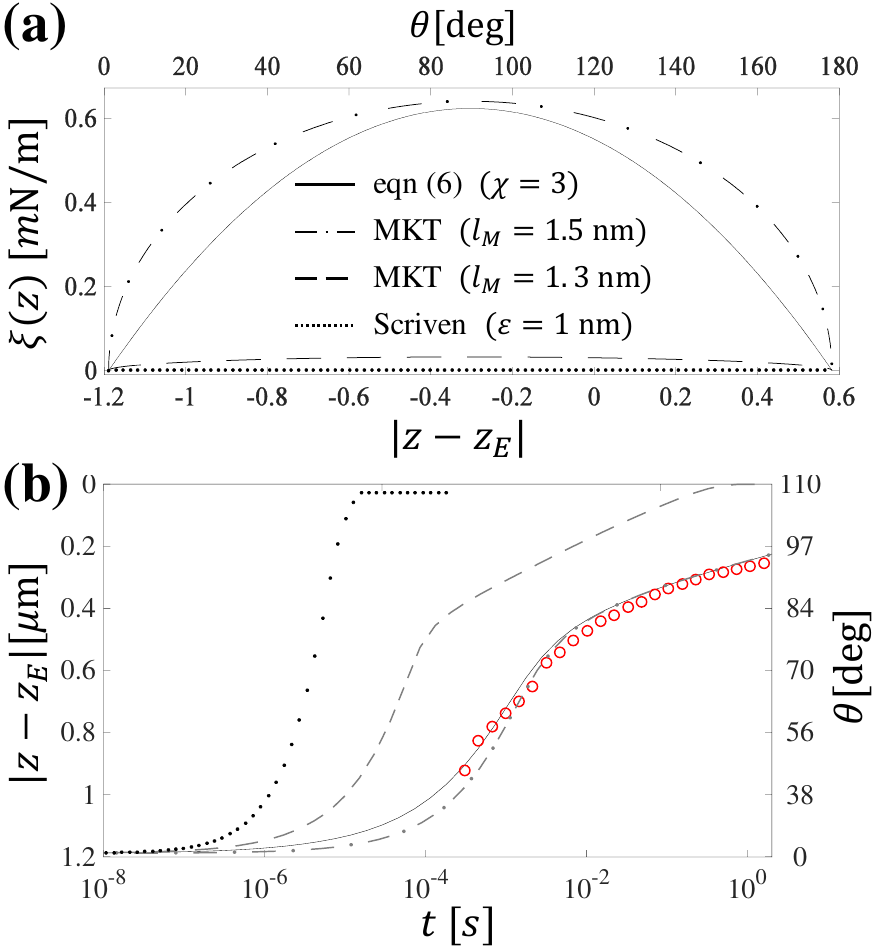}
\vskip -5pt
\caption{Dissipative forces and adsorption dynamics.
(a) Friction coefficient $\xi(z)$ versus particle separation from equilibrium $|z-z_E|$.
(b) Separation from equilibrium $|z-z_E|$ versus time adopting friction coefficients reported in (a).
Results correspond to the carboxyl functionalized particle reported in Table~\ref{tb:parameters}.
Solid lines: Friction coefficient derived via Green--Kubo relations (eqn~(\ref{eq:xis})) with $\chi=3$.
Dashed-dotted lines: MKT friction coefficient for $\lambda_M=$~1.5 nm.
Dashed lines: MKT friction coefficient for $\lambda_M=$~1.3 nm.
Dotted lines: Scriven's hydrodynamic models with molecular cutoff parameter $\varepsilon=$~1 nm.
}
\label{fig4}
\end{figure}
%%%%%%%%%%%%%%%%%%%%%%%%%%%%%%%%%%%%%%%%%%%%%%%%%%%%%%%%%%%%%%%%%%%%%%%%%%%%%%%%%%%%%%%%%%%%%%%%%%%%%%%%%%%%%%%%%%%%%%%%%%%%%%%%

To determine the dissipative force $F_d=-\xi \dot{z}$ employed in eqn~(\ref{eq:langevin}) we considered different contributions predicted by hydrodynamic models, Blake's MKT theory for contact line dynamics, and eqn~(\ref{eq:xis}) derived by invoking the fluctuation-dissipation theorem.  
For the case of hydrodynamic drag contributions, we found that models based on Stokes flow predict friction coefficients that are smaller than or comparable to those predicted by Scriven's corner flow model in the limit of physically 
meaningful values of the molecular cutoff length $\varepsilon$.\cite{huh1971hydrodynamic}
In Fig.~\ref{fig4}a we report the value of the position-dependent friction coefficient $\xi(z)$ predicted for the studied experimental conditions by three different models: 
(1) Scriven's hydrodynamic model for corner flow adopting a molecular cutoff length $\varepsilon=1$\,nm,\cite{huh1971hydrodynamic}
(2) MKT predictions\cite{blake2006physics} for two adsorption site sizes $\lambda_M=$~1.3\,nm and $\lambda_M=$~1.5\,nm, 
and (3) the expression derived in eqn~(\ref{eq:xis}) with a correlation factor $\chi=3$.
For a carboxyl functionalized particle with $R=0.9~\mu$m and nanoscale defect area $A_d=$~12\,nm$^2$ [see Table~1], eqn~(\ref{eq:xis}) predicts a maximum friction coefficient $\xi_s(z=0)\simeq$~6.2$\times 10^{-4}$~N~s/m,
which is within the magnitude range required for agreement between the adsorption dynamics modeled by eqn~(\ref{eq:langevin}) and experimental observations.\cite{kaz2012,wang2013}
As showed in Figs.~\ref{fig4}a--b, large dissipative forces are required to describe the experimentally observed trajectories. 
Such dissipative forces are predicted by the friction coefficient $\xi_s$ in 
eqn~(\ref{eq:xis}), which can be several orders of magnitude larger than friction coefficients due to hydrodynamic and molecular kinetic effects.   
Employing MKT predictions with an adsorption site of size $\lambda_{M}=1.5$\,nm results in a friction coefficient $\xi$ that is sufficiently large to produce a reasonable (but not optimal) fit to the experimental data [cf Fig.~\ref{fig4}b].
However, the energy barrier $\Delta{\cal F}=19.2~k_B T$ employed to describe the logarithmic relaxation rate and estimate the crossover point via eqn~(\ref{eq:crossover}) is significantly larger than the work of adsorption $W_a=13.9~k_B T$ corresponding to $\lambda_{M}=1.5$\,nm.
Moreover, the use of molecular adsorption sites with dimensions $\lambda_{M}\ge 1$\,nm is beyond the original premises of Blake's MKT.
As indicated by Blake and co-workers,\cite{blake2011dynamics}
modeling the effects of nano- and mesoscale surface defects is likely to require the use of alternative models based on Kramers theory.\cite{colosqui2013,razavi2014}   
Notably, for the range of defect sizes $A_d=$~9.2--85.7\,nm$^2$ employed to fit the logarithmic relaxation rates of different particles, the friction coefficient $\xi_s$ (eqn~(\ref{eq:xis})) derived from the fluctuation-dissipation theorem accounts quantitatively for the observed adsorption dynamics.
%

%%%%%%%%%%%%%%%%%%%%%%%%%%%%%%%%%%%%%%%%%%%%%%%%%%%%%%%%%%%%%5
%
% CONCLUSIONS
%
%%%%%%%%%%%%%%%%%%%%%%%%%%%%%%%%%%%%%%%%%%%%%%%%%%%%%%%%%%%%%%

\section{Conclusions}
We have proposed a Langevin dynamics approach for the adsorption of single colloidal particles that considers deterministic spatial fluctuations of the free energy induced by nanoscale surface defects and random thermal fluctuations of the particle position and local contact angle with the liquid interface.
The proposed Langevin model is able to describe the entire adsorption dynamics of single microparticles with different surface functional groups that has been reported in previous experimental work.
The adsorption dynamics exhibits two remarkably different regimes: (i) capillary-driven adsorption dominated by reduction of surface energy, and (ii) thermally activated adsorption governed by random transitions between multiple metastable states. 
The capillary-driven regimes are observed during the initial stages of the adsorption process and are damped by friction forces that are much larger than predicted by previously available drag models.
In our model, the thermal motion of the interface produces off-plane fluctuating surface forces that have zero mean and finite variance and are uncorrelated with the off-plane translational motion of the particle.  
The energy injected into the particle translation by fluctuating surfaces forces must be dissipated by a large damping force in order to satisfy the fluctuation-dissipation relation for a system in thermal equilibrium. 
By means of the Green-Kubo formalism we have derived a new analytical expression that predicts dissipative forces with the magnitude and functional dependence on the particle position that are required to describe the experimentally observed dynamics of adsorption.  
The crossover to slow thermally activated regimes observed in experiments and simulations can be predicted by a simple analytical criterion (eqn~(\ref{eq:crossover})) based on the emergence of local energy minima when thermodynamic equilibrium is approached.

The proposed Langevin model allows us to elucidate novel aspects of the dynamics of adsorption of single colloidal microspheres at liquid-fluid interfaces and make a few useful predictions for further experimental validation.
For example, invoking the crossover criterion in eqn~(\ref{eq:crossover}) the characteristic defect area 
$A_d=\Delta{\cal F}/(\gamma\pi \alpha |z_C/R-\cos_E|)$, where $\alpha=$~0.8--0.9, could be estimated from the experimentally observed crossover position $z_C$, once the energy barrier $\Delta{\cal F}$ is determined by fitting the logarithmic regime. 
In the absence of nanoscale characterization techniques to determine the shape and characteristic area $A_d$ of surface defects, the proposed analytical model along with experimental observations can serve as a probe of such nanoscale features. 
A better understanding of the adsorption dynamics, as provided by the models in
this work, will help in determining fundamental properties of particles at interfaces, such as their equilibrium and non-equilibrium binding energy and contact angle.

We thank D.M. Kaz, J. Koplik, R. McGorty, and J.M. Morris for useful discussions. 
AW and VNM acknowledge support from the National Science Foundation under grant no. DMR-1306410 and from the Harvard MRSEC, supported by NSF grant no. DMR-1420570. 
AMR and CEC acknowledge support from the SEED
Grant Program by The Office of Brookhaven National Laboratory (BNL) Affairs at Stony Brook University (SBU).
This work employed computational resources from the Institute for Advanced Computational Science at SBU and the BNL Center for Functional Nanomaterials supported by the U.S. DOE under Contract No. DE-SC0012704.

\providecommand*{\mcitethebibliography}{\thebibliography}
\csname @ifundefined\endcsname{endmcitethebibliography}
{\let\endmcitethebibliography\endthebibliography}{}

\end{document}